\documentclass[aps,preprintnumbers,eqsecnum,amsmath,amssymb,nofootinbib]{revtex4}  
\usepackage{graphicx} 
\usepackage{bm} 
\usepackage{slashed}
\usepackage{color} 

\renewcommand\theequation{\arabic{section}.\arabic{equation}}
\begin{document}
\title{Constraining $\lambda_{B_s}$ by $B_s\to \gamma^*$ and $B_s\to \phi$ form factors}
\author{Mikhail A.~Ivanov$^{a}$, Dmitri Melikhov$^{a,b,c}$, and Silvano Simula$^{d}$}
\affiliation{
$^a$Joint Institute for Nuclear Research, 141980 Dubna, Russia\\
$^b$D.~V.~Skobeltsyn Institute of Nuclear Physics, M.~V.~Lomonosov Moscow State University, 119991, Moscow, Russia\\
$^c$Faculty of Physics, University of Vienna, Boltzmanngasse 5, A-1090 Vienna, Austria\\
$^d$INFN, Roma Tre, Via della Vasca Navale 84, I-00146 Rome, Italy}
\begin{abstract}
  We calculate the form factors $F_{V}(q^2,q'^2)$ and $F_{TV}(q^2,q'^2)$ describing the $B_s\to \gamma^*$ transition
  induced by the vector and tensor weak currents 
  in a broad range of values of $q^2$ and $q'^2$ far below the quark thresholds in both $q^2$ and $q'^2$ channels. 
  These form factors are calculated via the distribution amplitudes of the $B_s$-meson.  
  We then interpolate the obtained results by a formula that contains pole at $q'^2=M_\phi^2$ and
  extract the residue which gives the $B_s\to \phi$ transition form factors $V(q^2)$ and 
  $T_1(q^2)$. In this way we obtain theoretical predictions for these form factors 
  without invoking quark-hadron duality and QCD sum rules.
  Furthermore, we calculate the relationship between $V(0)$ and $T_1(0)$ and the parameter $\lambda_{B_s}(\mu)$,
  the inverse moment of the $B_s$-meson distribution amplitude. 
  Using the available predictions for $V(0)$ and $T_1(0)$ coming from approaches not referring
  to the $B_s$-meson distribution amplitudes, we obtain the estimate
  $\lambda_{B_s}(\mu\simeq m_b)=(0.62\pm 0.10)$ GeV. 
\end{abstract}
\date{\today}
\maketitle
\normalsize
 
\section{Introduction}
\label{Sec:introduction}
The inverse momentum of the distribution amplitude (2DA) $\phi_+$ of $B_{(s)}$-meson, $\lambda_{B_{(s)}}$,
is an interesting parameter for the phenomenology of weak $B_{(s)}$ decays as it enters a number
of observables related to semileptonic and multileptonic
decays of $B_{(s)}$-mesons (see e.g. \cite{korchemsky2000,braun2017,beneke2020,bbm2023,bbm2024,wang2023}). 
In spite of its importance, this parameter is presently not known with high accuracy and
theoretical predictions for this parameter vary in a rather broad range, namely
$\lambda_B(1 \mbox{ GeV})=0.3\div 0.7$ GeV \cite{kou,braun2004,beneke2011,wang2016,zwicky2021,im2022},
whereas the ratio $\lambda_{B_s}/\lambda_B$ is known with a better accuracy,
$\lambda_{B_s}/\lambda_B=1.19\pm 0.14$ \cite{khodjamirian2020}. 

In this paper, we discuss a new way of constraining $\lambda_{B_s}$.

The first step is the calculation of the form factors $F_{V}(q^2,q'^2)$ and $F_{TV}(q^2,q'^2)$
\cite{mk2003,mnk2018} describing the $B_s\to \gamma^*$ transition (see Fig.~1) in a broad
range of values of $q^2$ and $q'^2$ far below quark thresholds via
the 2DAs of the $B_s$-meson. For the relevant 2DAs we make use of the local duality (LD) model of \cite{braun2017}.
 
Then, we interpolate these results by an analytic formula which contains a number of poles
corresponding to vector meson resonances with quantum numbers appropriate to $q^2$ and $q'^2$
channels. This formula is based on a dispersion representation for $F_{V,TV}(q^2,q'^2)$ in $q'^2$
with one subtraction \cite{mnp2004} and the spectral density of this representation
is saturated by two resonances - the known lowest resonances $\phi$ and an effective vector meson pole with an unknown mass.
The numerical parameters in this analytic formula are obtained by fitting the results of our calculations
in a broad range of values of $q^2$ and $q'^2$ far below the quark thresholds in $q^2$ and $q'^2$ channels. 
The analytic formula contains a pole at $q'^2=M_\phi^2$; the residue of this pole
provides the $B_s\to \phi$ form factor $V(q^2)$ in the case of $F_V$ and $T_1(q^2)$ in the case of $F_{TV}$. 
Noteworthy, our estimate does not employ quark-hadron duality and QCD sum rules and is therefore free
of the intrinsic uncertainties of the method of QCD sum rules. Our approach thus provides an interesting
complementary method to the widely used method of QCD sum rules. 
 
For comparison, we present in Appendix \ref{AppendixA} the application of the standard QCD sum rule machinery
\cite{braun1994} in order to calculate $T_1(0)$ using the $B_s$-meson 2DAs \cite{offen2007}. 
We recall that QCD sum rules calculate the residue from the very
same Green function which determines the form factor $F_{TV}(q^2,q'^2)$. However, the QCD sum-rule extraction contains
auxiliary parameters (Borel parameter and effective threshold $s_{\rm eff}$) which are not fixed by any rigorous criteria, leading to
uncertainties that are difficult to control. Still we observe that using some commonly adopted ways of
fixing these auxiliary parameters leads to a good agreement between the estimates for $T_1(0)$ 
obtained by a direct calculation of the correlation function and extrapolation to the pole
and by a direct evaluation of the residue by a quark-hadron duality Ansatz.

The approach formulated above is based on using the representations for the form factors
$F_{V,TV}(q^2,q'^2)$ up to leading-order in the strong coupling constant $\alpha_s$. 
Making use of the known radiative corrections for the form factor $F_{V}(q^2,0)$ \cite{beneke2011},
we give arguments that for the form factors $F_{V,TV}(q^2,q'^2)$ the radiative corrections evaluated at the scale $\mu=m_b$
are at the level of less than 0.5\% in the broad range of values of $q^2$ and $q'^2$ used in our analysis.
This property means that the leading-order expression for $F_{V,TV}(q^2,q'^2)$ expressed via the
2DAs at the scale $\mu=m_b$ should give a good estimate for the physical form factors $F_{V}(q^2,q'^2)$ and $F_{TV}(q^2,q'^2|\mu=m_b)$. 

We then obtain the dependence of the $B_s\to\phi$ form factors $V(0)$ and $T_1(0)$ 
on $\lambda_{B_s}(\mu=m_b)$. Combining our results with the predictions for $V(0)$ and $T_1(0)$ obtained by
approaches not referring to 2DAs of the $B_s$-meson, we extract 
\begin{eqnarray}
\lambda_{B_s}(\mu\simeq m_b)=(0.62\pm 0.10) \mbox{ GeV}. 
\end{eqnarray}

\section{The form factors $F_{V,TV}(q^2,q'^2)$ and the $B_s\to\phi$ form factors $V(q^2)$ and $T_1(q^2)$}
This Section provides details of the analytic calculation of $F_{V,TV}(q^2,q'^2)$ and the corresponding numerical results. 
\subsection{Analytic results for $F_{V,TV}(q^2,q'^2)$ at leading order in $\alpha_s$} 
\begin{figure}[b!]
\begin{center}
\includegraphics[height=5cm]{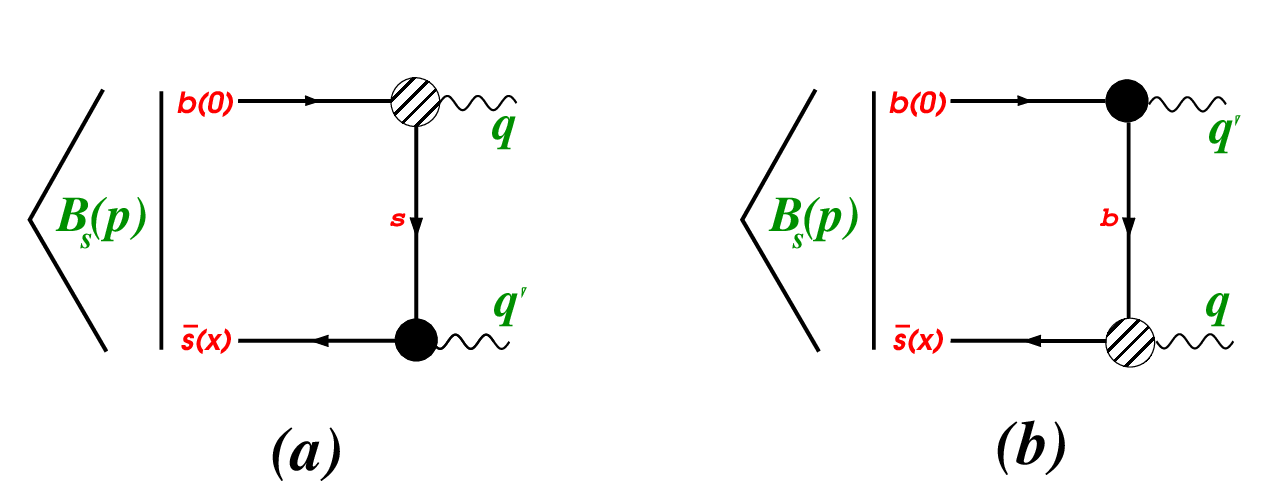} 
  \caption{(a) Diagram corresponding to the form factor $F^{(s)}_{TV}(q^2,q'^2)$ containing $\phi$-meson
 pole in the $q'^2$-channel. (b) Diagram corresponding to parametrically suppressed form factor $F^{(b)}_{TV}(q^2,q'^2)$. 
 The dashed blob stands for the tensor current $\bar s \sigma_{\mu\nu}b$ and the solid blob
    stands for the electromagnetic current, $j_\alpha^{\rm e.m.}=Q_s \bar s \gamma_\alpha s+Q_b \bar b \gamma_\alpha b$. }
\label{Fig:Bs2gamma}
\end{center}
\end{figure}
Let us consider the transition amplitudes involving electromagnetic
and weak currents \cite{mk2003,mnk2018} (Fig.~\ref{Fig:Bs2gamma})\footnote{
Our notations and conventions are: 
$\gamma^5=i\gamma^0\gamma^1\gamma^2\gamma^3$, 
$\sigma_{\mu\nu}=\frac{i}{2}[\gamma_{\mu},\gamma_{\nu}]$, 
$\epsilon^{0123}=-1$, $\epsilon_{abcd}\equiv
\epsilon_{\alpha\beta\mu\nu}a^\alpha b^\beta c^\mu d^\nu$, 
$e=\sqrt{4\pi\alpha_{\rm em}}$.}:  
\begin{eqnarray}
i\int dx e^{i q' x} 
\langle 0|\left\{T j_\alpha^{\rm e.m.}(x), \bar s \gamma_{\mu} q^\nu b(0)\right\}|\bar B_s(p)\rangle
&=&ie\,\epsilon_{\mu \alpha q q'}\frac{F_{V}(q^2,q'^2)}{M_{B_s}},\nonumber \\
 \label{tensor_penguin}
i\int dx e^{i q' x} 
\langle 0|\left\{T j_\alpha^{\rm e.m.}(x), \bar s \sigma_{\mu\nu} q^\nu b(0)\right\}|\bar B_s(p)\rangle
&=&ie\,\epsilon_{\mu \alpha q q'}F_{TV}(q^2,q'^2). 
\end{eqnarray}
The form factors $F_{i}(q^2,q'^2)$, [$i=V,TV$] in Eq.~(\ref{tensor_penguin}) contains two contributions,
$F_{i}=F^{(s)}_{i}+F^{(b)}_{i}$ corresponding to the diagrams in Fig.~\ref{Fig:Bs2gamma}.
We are interested in $F^{(s)}_{i}$ which contains a pole at $q'^2=M^2_\phi$ and neglect 
the form factor $F^{(b)}_{i}$ parametrically suppressed by
$1/m_b$ compared to  $F^{(s)}_{i}$. So, hereafter $F_{i}$ denotes $F^{(s)}_{i}$.

The form factors $F_{i}$  may be calculated via the $B$-meson 2DAs using HQET formula
(see e.g. \cite{offen2007})
\begin{eqnarray}
  \label{2BS}
  \langle 0|\bar s(x)\Gamma b(0)|\bar B_s(p)\rangle=-\frac{if_{B_s} M_{B_s}}{4}\int d\xi e^{-i\xi  p x}
  \,{\rm Tr}\bigg\{ \gamma_5 \Gamma (1+\slashed{v})\bigg[\phi_+(\xi)-\frac{\phi_+(\xi)-\phi_-(\xi)}{2 vx}\slashed{x}\bigg]\bigg\},
  \quad p_\mu=M_{B_s}v_\mu.
\end{eqnarray}
Omitting all contributions at order $O(m_s/M_{B_s})$ in the numerator, one gets
\begin{eqnarray}
\label{FV}
F_{V}(q^2,q'^2)&=&-Q_s f_{B_s} M_{B_s}
\int d\xi
\bigg[
\frac{\phi_+(\xi)}{m_s^2+\xi(1-\xi)M_{B_s}^2-q^2 \xi -q'^2(1-\xi)}
+
\frac{2 m_s M_{B_s} \bar \Phi(\xi)}{(m_s^2+\xi(1-\xi)M_{B_s}^2-q^2 \xi -q'^2(1-\xi))^2}\bigg],
\nonumber
\\
\label{FT}
F_{TV}(q^2,q'^2)&=&-Q_s f_{B_s} M_{B_s}
\int d\xi \frac{\phi_+(\xi)(1-\xi)+\bar\Phi(\xi)}{m_s^2+\xi(1-\xi)M_{B_s}^2-q^2 \xi -q'^2(1-\xi)}
\end{eqnarray}
with
\begin{eqnarray}
\bar \Phi(\xi)=\int\limits_0^\xi d\xi' \big[\phi_{+}(\xi')-\phi_{-}(\xi')\big]. 
\end{eqnarray}
The kinematical singularity at $vx\to 0$ in Eq.~(\ref{2BS}) cannot be a singularity of the
physical amplitude, so that  the primitive $\bar \Phi(\xi)$ should vanishe at the boundaries of the 2DA support region.
Then the integration by parts in $\xi$, necessary to handle the $1/vx$ term, does not contain any nonzero surface term. 

For the 2DAs we use a set referred to as Model IIB in \cite{braun2017} (our $\xi$ and $\omega_0$ are dimensionless): 
\begin{eqnarray}
\label{2DAsa}
\phi_+(\xi)&=&\frac{5}{8\omega_0^5}\xi (2\omega_0-\xi)^3\theta(2\omega_0-\xi),\\
\label{2DAsb}
\phi_-(\xi)&=&\frac{5}{192\omega_0^5}(2\omega_0-\xi)^2\bigg[
  6(2\omega_0-\xi)^2-\frac{7(\lambda_E^2-\lambda_H^2)}{M_{B_s}^2\omega_0^2}
 (15\xi^2-20\omega_0\xi+4\omega_0^2) \bigg]
\theta(2\omega_0-\xi),
\end{eqnarray}
where
\begin{eqnarray}
\label{omega0}
\omega_0=\frac{5}{2}\frac{\lambda_{B_s}}{M_{B_s}}. 
\end{eqnarray}
For numerical estimates we use $M_{B_s}$=5.367 GeV and $f_{B_s}$=0.230 GeV.  
The parameters $\lambda^2_{E,H}\simeq (0.1-0.2)$ GeV$^2$ are related to certain matrix elements
of antiquark-quark-gluon operators \cite{braun2017}. The contribution of the term proportional to
$(\lambda_E^2-\lambda_H^2)$ in Eq.~(\ref{2DAsb}) for the form factors $F_{V,TV}$ turns out to be numerically
negligible and may be safely omitted. For a detailed analysis of the analytic properties of the form factor
(\ref{FT}) for the case of the polynomial 2DAs we refer to \cite{ims2020}. 

\subsection{Numerical results for $F_{V}(q^2,q'^2)$ and $F_{TV}(q^2,q'^2)$}
We now obtain the form factors $F_{i}(q^2,q'^2)$, [$i=V,TV$] in a broad range of values of $q^2$ and $q'^2$ by the following procedure:

\noindent (i) $F_{i}(q^2,q'^2)$ is calculated using Eq.~(\ref{FT}) 
with 2DAs $\phi_{\pm}$ given in Eqs.~(\ref{2DAsa}) and (\ref{2DAsb}) for $\lambda_{B_s}=0.6$ GeV.

\noindent (ii) We interpolate the numerical results in the range $0<q^2({\rm GeV^2})<6$ and $-10<q'^2({\rm GeV^2})<-0.6$
(i.e., far below the quark thresholds located at $q^2=(m_b+m_s)^2$ and $q'^2=4m_s^2$) using a simple analytic fit formula 
\begin{eqnarray}
\label{fitFT}
F(y_1,y_2)&=&
f_{0}\left(1+a_{0}\frac{y_1}{1-y_1}\right)+
R_\phi F_{0}\left(1+a_{1}\frac{y_1}{1-y_1}\right)\frac{y_2}{1-y_2}+
R_\phi F_{1}\left(1+a_{2}\frac{y_1}{1-y_1}\right)\frac{y_2}{1-y_2/r^2},\nonumber\\
  && 
  y_1\equiv q^2/M_{B^*_s}^2, \; y_2\equiv q'^2/M^2_{\phi},
\end{eqnarray}
where
$R_\phi=-\frac{2 f^{\rm e.m.}_\phi}{M_\phi}, f^{\rm e.m.}_\phi=Q_sf_\phi, Q_s=-1/3, f_\phi=0.224 \mbox{ GeV}, M_\phi=1.020  \mbox{ GeV},
M_{B_s^*}=5.415\mbox{ GeV}$. The analytic formula (\ref{fitFT}) reflects the correct location
of the lowest physical poles at $q^2=M_{B^*_s}^2$ and $q'^2=M^2_{\phi}$.

The $B_s\to\phi$ transition form factors at $q^2=0$ are related to the residues in the pole at $q'^2=M_\phi^2$
such that $V(0)=\frac{M_{B_s}+M_\phi}{M_{B_s}}F_{0,V}$ and $T_1(0)=F_{0,TV}$. 

The formula (\ref{fitFT}) is suggested by the following theoretical considerations:
 
\noindent
$\bullet$ We consider a dispersion representation in $q'^2$ with one subtraction and saturate its 
spectral density by the sum of two poles, a physical pole 
at $q'^2=M_\phi^2$ and an effective heavier pole at $q'^2=M_\phi^2/c$, $c<1$. 

\noindent
$\bullet$ 
The coefficients in this $q'^2$-dispersion representation are functions of $q^2$.
For each of these functions we again write down a spectral representation in $q^2$
with one subtraction and saturate the spectral density by a lowest vector meson $B_s^*$. 
One may include also the effective heavier vector pole in the $q^2$-channel, but this makes no difference
as already the $B_s^*$ lies far above the $q^2$-region of our interest. 

\vspace{0.5cm}

\begin{figure}[h!]
\begin{center}
\begin{tabular}{cc}   
  \includegraphics[height=5cm]{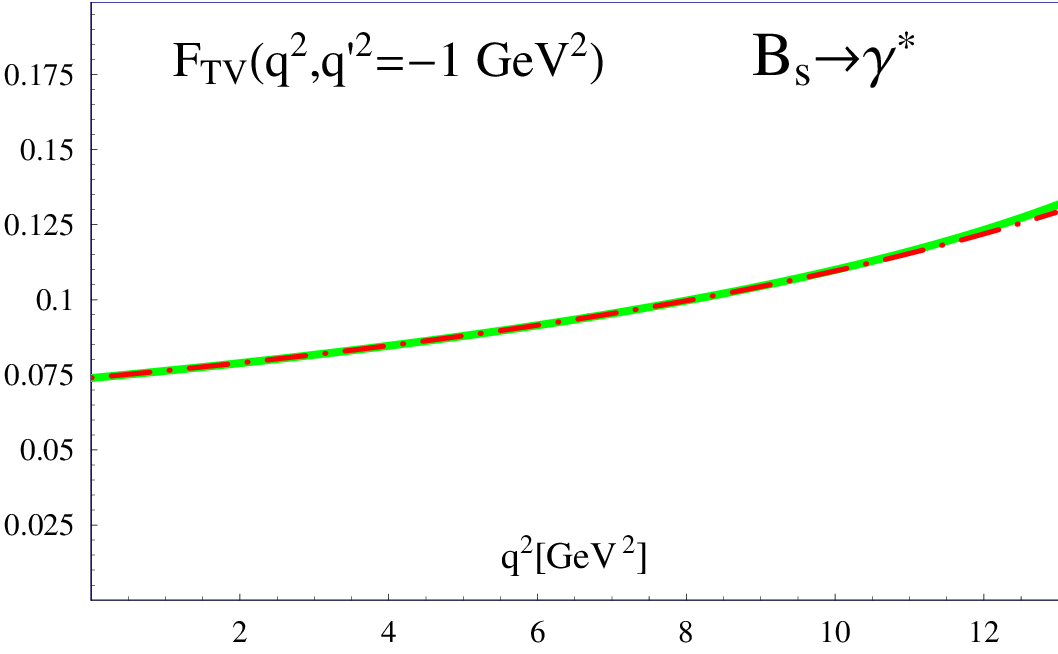} &
  \includegraphics[height=5cm]{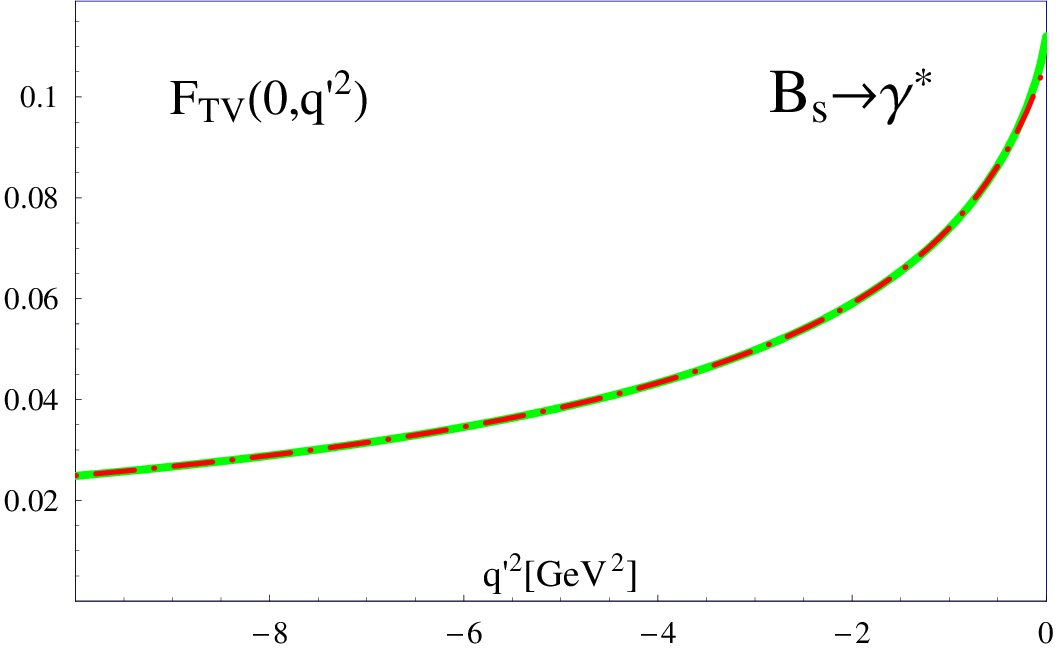}\\
  (a)    &     (b)\\
  \end{tabular}
\caption{\label{Plots:FTV}
  Solid (green) line is the results of direct calculations of the form factor $F_{TV}(q^2,q'^2)$ for $\lambda_{B_s}=0.6$ GeV.
  Dashed (red) line is the results of the double fit to $F_{TV}(q^2,q'^2)$ in the range $-10<q'^2({\rm GeV}^2)<-0.6$ and
  $0<q^2({\rm GeV}^2)< 6$.   
\label{Plot:H}}
\end{center}
\end{figure}

Table \ref{Table:fitFT} gives the fit parameters obtained by this interpolation procedure.
\begin{table}[b!]
\centering
\caption{Parameters of the fit formula Eq.~(\ref{fitFT}) obtained by the interpolation of our numerical results for $F_{V,TV}$
  calculated for $\lambda_{B_s}=0.6$ GeV. The corresponding $B_s\to \phi$ form factors are $V(0)=0.35$ and $T_1(0)=0.33$.}
  \label{Table:fitFT}
\begin{tabular}{|l|r|r|r|r|r|r|r|}
  \hline
         & $f_{0}$  &  $a_{0}$   &  $F_{0}$  & $a_{1}$  & $F_{1}$ & $a_{2 }$    & $r$  \\
  \hline
$F_{V}$   & 0.13   &  $1.36$  &  $0.29$  & $4.57$  & $0.126$ & $-0.25$     &  $2.17$  \\  
  \hline
$F_{TV}$  & 0.11   &  $1.23$  &  $0.33$  & $2.62$  & $0.085$ & $0.26$      &  $2.20$  \\  
  \hline
\end{tabular}
\end{table}
The fit formula (\ref{fitFT}) interpolates the results of calculating
$F_{V,TV}(q^2,q'^2)$ with an accuracy better than 1\% in the full adopted range of values of $q^2$ and $q'^2$. 
Fig.~\ref{Plots:FTV} illustrates the quality of reproducing the calculation results by the fit formula.

One can reach a much better accuracy (at a per mille level) if one proceeds in a bit different way: namely,
performs a fit to the calculated $F_{TV}(q^2,q'^2)$ in the single variable $q'^2$ in the range
$-10 <q'^2 \;({\rm GeV}^2) <-0.6$ for any fixed value of $q^2$ far below the quark threshold located at $(m_b+m_s)^2$.
We follow this route when calculating the dependence of $T_1(0)$ and $V(0)$ on $\lambda_{B_s}$ (see below). 

As suggested by Eq.~(\ref{fitFT}), in the region $0<q^2({\rm GeV}^2)<6$ the results of our calculation may
be well approximated by a simple monopole formula $F(q^2)=F_{0}\left(1+a_1 y_1/(1-y_1)\right)$ with $y_1=q^2/M^2_{B_s^*}$.

\subsection{Radiative corrections and constraints on $\lambda_{B_s}(\mu)$}
It is time to recall that the analysis of the previous Section does not include the radiative corrections, 
whereas for the form factors of the tensor current, $F_{TV}(q^2,q'^2)$ and $T_1(q^2)$ which depend on the scale $\mu$,
the radiative corrections are crucial for controlling this scale-dependence. Also for the form factors
$F_{V}(q^2,q'^2)$ and $V(q^2)$, the radiative corrections may change the dependence of these form factors on $\lambda_{B_s}(\mu)$. 

Two remarks are in order.

\begin{figure}[h!]
  \begin{center}
    \includegraphics[height=5cm]{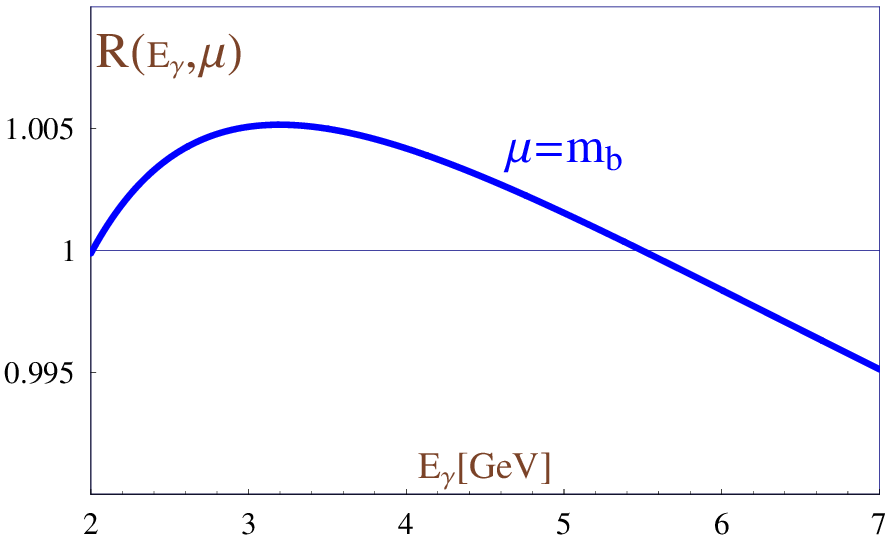}  
\caption{\label{Plots:RvsE}
The factor $R(E_\gamma,\mu=m_b)$ describing radiative corrections to the form factor $F_{V}(E_\gamma)$. }
\end{center}
\end{figure}

\noindent
$\bullet$ The radiative corrections are known for the case $F_{V}(q^2,q'^2=0)$ as the function of
$E_\gamma=(M_{B_s}^2-q^2)/(2M_{B_s})$, see 
\cite{beneke2011} and Appendix \ref{AppendixB}. The corresponding factor $R(E_\gamma,\mu)$ is shown in Fig.~\ref{Plots:RvsE} for $\mu=m_b$.
As seen from this plot, the correction evaluated at $\mu=m_b$ is negligible in the range $2<E_\gamma({\rm GeV}^2<7$.
Recall, however, that we have a different situation: namely, $q^2\ne 0$ and $q'^2 < 0$. We now give arguments
that the results for $F_{V}(q^2,q'^2=0)$ may be used to estimate the radiative corrections for $F_{V}(q^2,q'^2)$ at $q'^2<0$. 

Let us introduce the ``equivalent'' energy $\bar E_\gamma(q^2,q'^2)$ by the following
expression\footnote{Note that $\bar E_\gamma$ is different from the virtual photon energy for the process $B_s\to\gamma(q)\gamma(q')$
which is $E_\gamma(q^2,q'^2)=(M_{B_s}^2+q'^2-q^2)/(2M_{B_s})$.}: 
\begin{eqnarray}
m_s^2+M_{B_s}^2\bar \xi(1-\bar\xi)-q^2\bar\xi-q'^2(1-\bar\xi)\equiv 2M_{B_s}\bar\xi\bar E_\gamma(q^2,q'^2), 
\end{eqnarray}
such that the denominator in Eq.~(\ref{FV}) at $\xi=\bar\xi$ gives $2M_{B_s}\bar E_\gamma\bar\xi$.
Obviously, $\bar E_\gamma\to E_\gamma$ for $m_s\to 0$, $q'^2\to 0$, and $\bar\xi\to 0$. 

The behaviour of the form factor and of the radiative corrections are related to the behaviour of the
denominator of Eq.~(\ref{FV}). So, we expect that the radiative correction in $F_V(q^2,q'^2)$ may be estimated 
using the function $R(\bar E_\gamma,\mu)$ evaluated for $\bar\xi$, the value at which $\phi_+(\xi)$ reaches its maximum.
For the 2DAs considered here one obtains $\bar\xi\simeq 0.2$, and it is easy to check that for the range
used in our analysis, namely 
$-10 < q'^2({\rm GeV}^2)<0$ and $0 <q^2({\rm GeV}^2)< 6$, $\bar E_\gamma$ varies in the range
$2<\bar E_\gamma({\rm GeV}^2)<7$. 
Thus, for the analysis of the previous Section, the radiative corrections in the form factor $F_V$ are
at the level of less than 1\% and may be neglected, if one expresses the leading-order contribution to the form
factor $F_V(q^2,q'^2)$, Eq.~(\ref{FV}), via $\lambda_{B_s}(\mu=m_b)$. 

\vspace{0.5cm}

\noindent
$\bullet$
The radiative corrections for the form factor $F_{TV}(q^2,q'^2)$ have not yet been calculated for any
values of $q^2$ and $q'^2$. Again, since the behaviours of the form factors and of the radiative correction are determined by the
same denominator appearing in the analytic formulas for $F_{TV}(q^2,q'^2)$ and $F_{V}(q^2,q'^2)$, we {\it conjecture} that
the radiative corrections at $\mu=m_b$ in the considered range of values of $q^2$ and $q'^2$ may be neglected too.

Moreover, the form factors of the tensor current depend on the scale, 
in particular, one has $F_{TV}(0,0|\mu)$ and $T_1(0|\mu)$.
Our arguments for a negligible size of the radiative corrections at $\mu=m_b$ imply that the analysis
of the previous Section yields $F_{TV}(0,0|\mu=m_b)$ and $T_1(0|\mu=m_b)$ as soon as 
the leading-order contributions in $\alpha_s$, Eq.~(\ref{FV}), are expressed via $\lambda_{B_s}(\mu=m_b)$. 

\vspace{0.5cm}

We would like to emphasize that for our extraction of $\lambda_{B_S}(\mu)$ it is not crucial that the radiative corrections
vanish precisely at $\mu=m_b$; it is only important that these corrections vanish at $\mu \sim m_b$.
We shall see later (Fig.~\ref{Plots:lambdavsmu}) that the dependence of $\lambda_{B_S}(\mu)$ in the region $\mu \sim m_b$
is relatively flat and, therefore, the impact of the precise value of the scale at which the radiative corrections vanish,
has only a moderate impact on, e.g., the extracted value of $\lambda_{B_s}(\mu = m_b)$.

\begin{figure}[t!]
\begin{center}
  \begin{tabular}{cc}
 \includegraphics[height=4.5cm]{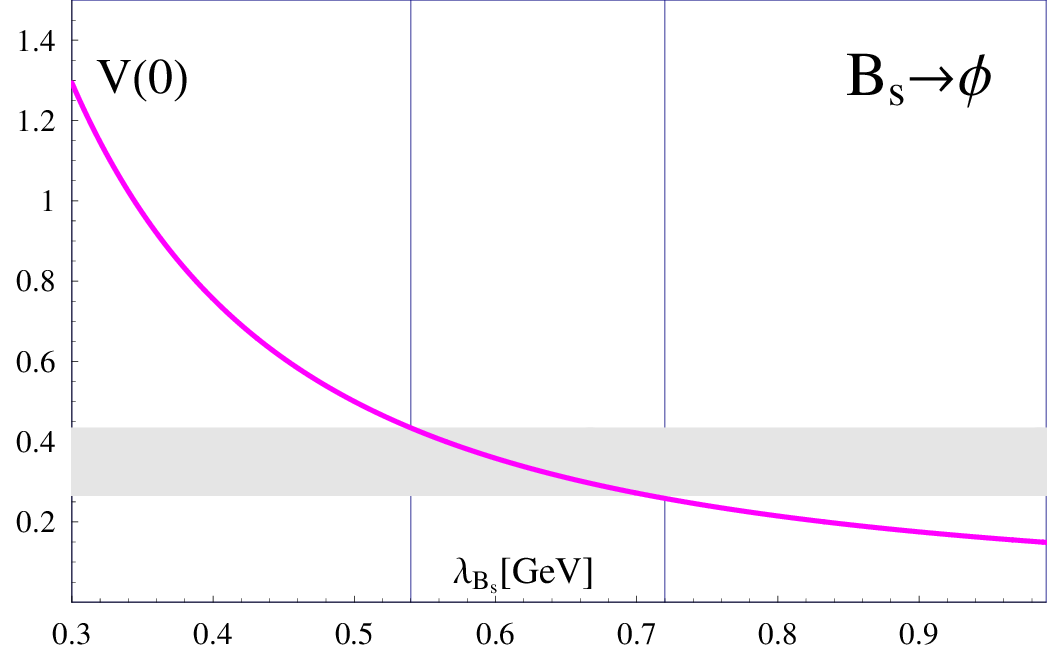}  &
 \includegraphics[height=4.5cm]{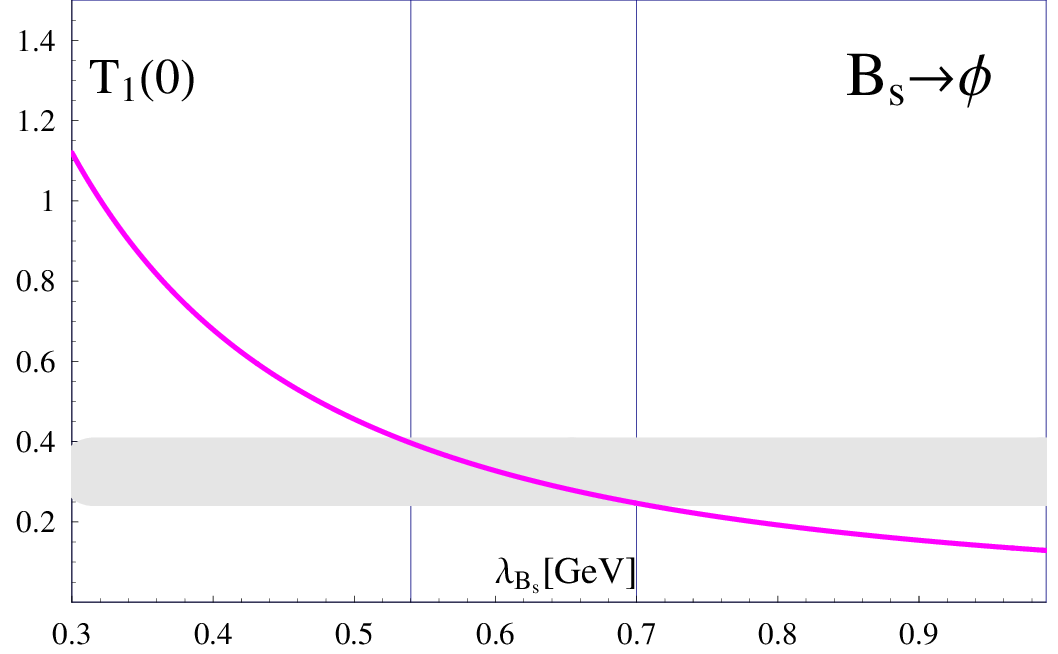} \\
     (a)  &   (b)\\
  \includegraphics[height=4.5cm]{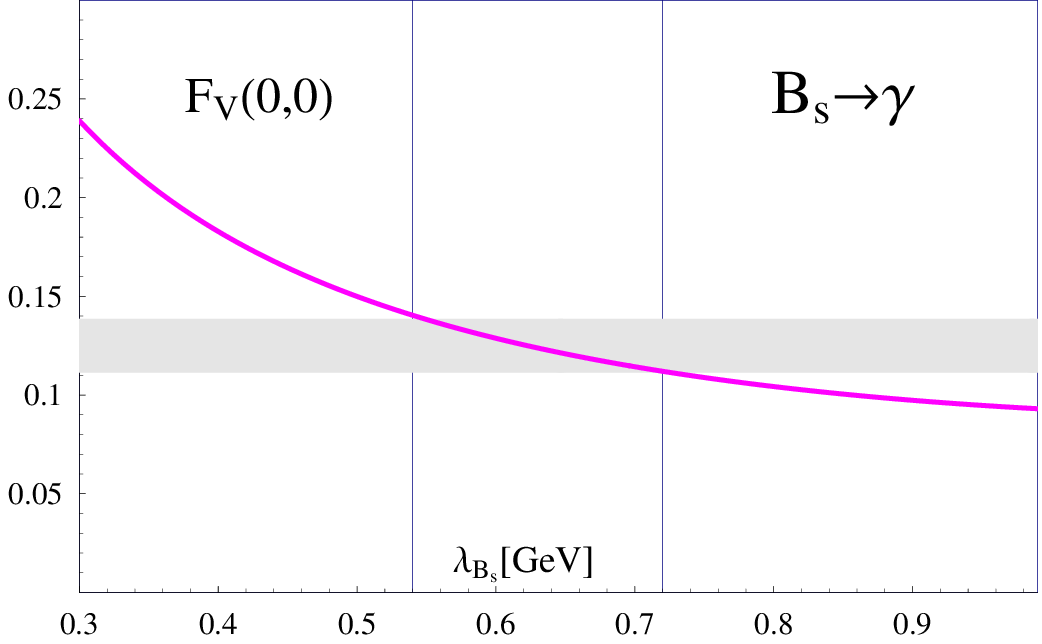}  &
  \includegraphics[height=4.5cm]{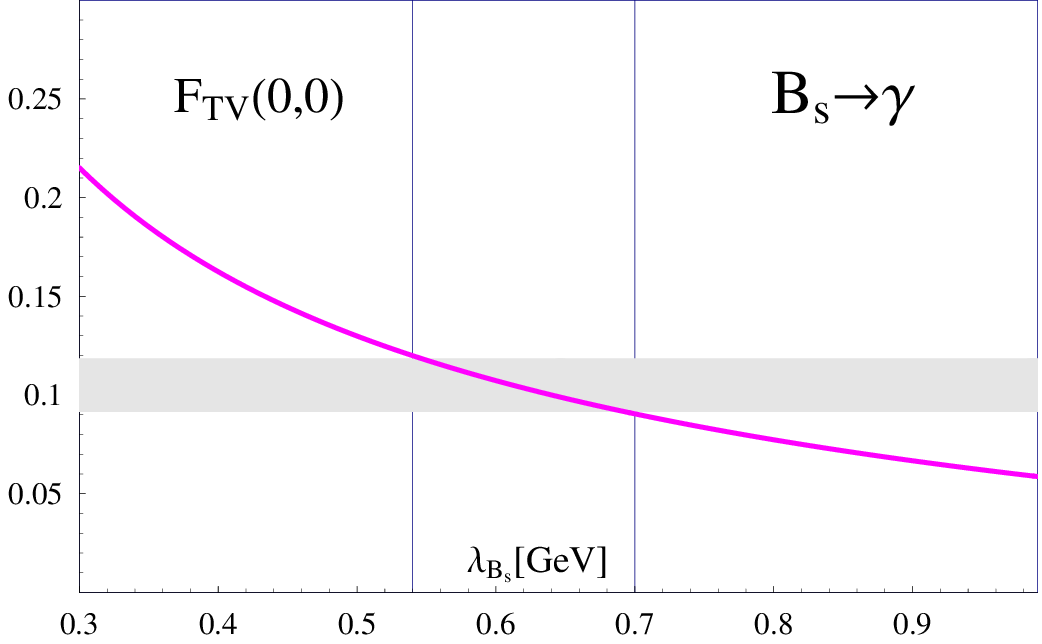} \\
  (c)  &   (d)\\
    \end{tabular}
  \caption{\label{Plots:FT}
(a) Form factor $V(0)$ calculated via the leading order diagram of Fig.~1~(a) 
    vs $\lambda_{B_s}=\lambda_{B_s}(m_b)$. The band covers the results for $V(0)$ in Table \ref{Table:T1}.
(b) Form factor $T_1(0)$ at the scale $\mu=m_b$ calculated via the leading order diagram of Fig.~1~(a) 
    vs $\lambda_{B_s}=\lambda_{B_s}(m_b)$. The band covers the results for $T_1(0)$ in Table \ref{Table:T1}.
(c) Form factor $F_{V}(0,0)$ vs $\lambda_{B_s}$.
    The band of values of $F_{V}(0,0)$ corresponds to the allowed band of $\lambda_{B_s}(m_b)$ from Plot.~(a)
(d) Form factor $F_{TV}(0,0)$ at the scale $\mu=m_b$ vs $\lambda_{B_s}$.
    The band of values of $F_{TV}(0,0)$ corresponds to the allowed band of $\lambda_{B_s}(m_b)$ from Plot.~(b).}
\end{center}
\end{figure}

\begin{figure}[h!]
\begin{center}
\begin{tabular}{cc}   
  \includegraphics[height=5cm]{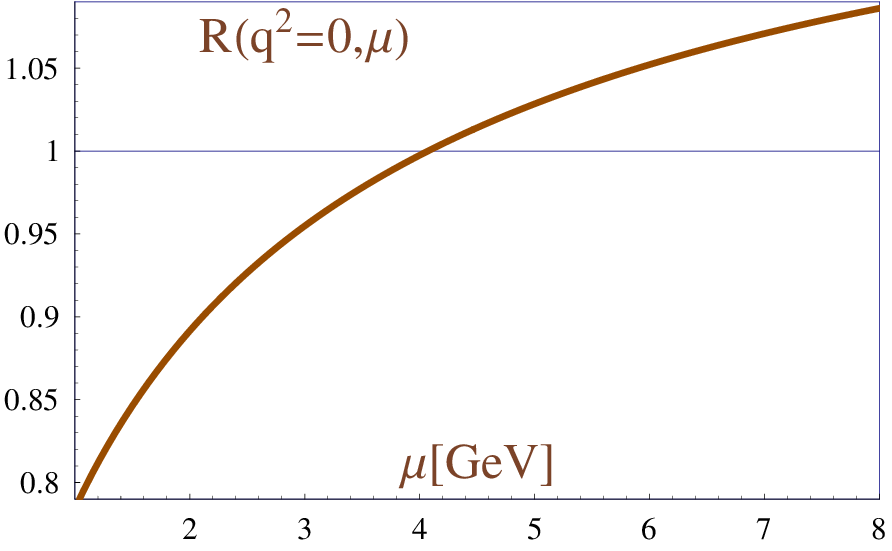} &   \includegraphics[height=5cm]{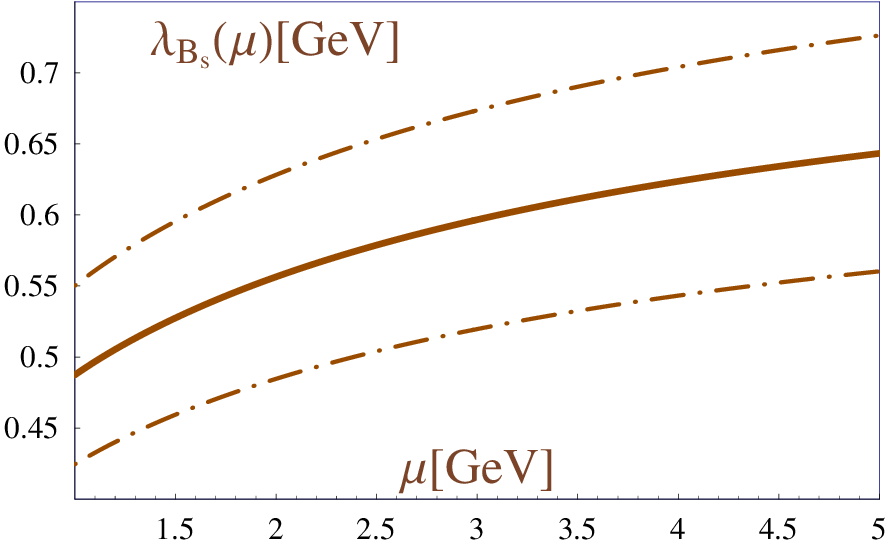}\\
  (a)  &   (b)\\
    \end{tabular}
\caption{\label{Plots:lambdavsmu}
  (a)  The dependence of the factor $R(E_\gamma=M_{B_s}/2,\mu)$ on $\mu$.
  (b)  The dependence of $\lambda_{B_s}(\mu)$ on $\mu$. Solid line describes 
  the evolution of the central value of the allowed interval of $\lambda_{B_s}(m_b)$ whereas the dash-dotted
  lines correspond to the lower and the upper boundaries of the allowed interval of $\lambda_{B_s}(m_b)$.}
\end{center}
\end{figure}

Fig.~\ref{Plots:FT} presents the results based on Eqs.~(\ref{FV}) showing strong correlations between
the value of $\lambda_{B_s}\equiv \lambda_{B_s}(m_b)$ 
and the form factor values $F_{V}(0,0)$, $V(0)$, $F_{TV}(0,0|\mu=m_b)$ and $T_1(0|\mu=m_b)$. 
In order to disentangle the appropriate value of $\lambda_{B_S}(m_b)$ we use the existing estimates of 
the $B_s\to\phi$ form factors $V(0)$ and $T_1(0|m_b)$, obtained using approaches not directly referring
to the 2DAs of $B_s$-meson.
\begin{table}[b!]
\centering 
\caption{Theoretical results for the $B_s\to\phi$ form factors $V(0)$ and $T_1(0|\mu=m_b)$
  and the corresponding $\lambda_{B_s}(m_b)$
obtained from Fig.~\ref{Plots:FT}.}
\label{Table:T1}
\begin{tabular}{|c|c|c|c|c|c|}
  \hline
                       & \cite{ms2000}  & \cite{ballbraun1998}  &   \cite{ballzwicky2005}  & \cite{lattice2014} & \cite{ivanov2016} \\
  \hline
  $V(0)$     &  $0.44\pm 0.05$  &   $0.43\pm 0.06$   &  $0.433\pm 0.035$   &  $0.24\pm 0.07$    &   $0.31\pm 0.03$   \\
$\lambda_{B_s}(\mu=m_b)[{\rm GeV}]$ &  0.54 &      0.55             &      0.55               &  0.72              &   0.65    \\  \hline
  $T_{1}(0|\,\mu=m_b)$  &  $0.38\pm 0.04$  &   $0.35\pm 0.05$   &  $0.349\pm 0.033$   &  $0.31\pm 0.02$    &   $0.27\pm 0.03$   \\
  $\lambda_{B_s}(\mu=m_b)[{\rm GeV}]$ &  0.54 &      0.58             &      0.58               &  0.62              &   0.7    \\
  \hline
\end{tabular}
\end{table}
Table \ref{Table:T1} shows a selection of such results obtained from quark models \cite{ms2000,ivanov2016}\footnote{The results obtained with
quark models do not include explicitly the radiative corrections and therefore do not control explicitly the scale-dependence
of the form factors of the tensor current. There are however arguments \cite{ms2000} that the tensor form factors obtained
in quark models correspond to the scale $\mu=m_b$.},
light-cone QCD sum rules based on the $\phi$-meson DAs \cite{ballbraun1998,ballzwicky2005},
and lattice QCD \cite{lattice2014}.

Conservatively, the existing results provide the allowed range
\begin{eqnarray}
V(0)=0.34\pm 0.10, \qquad T_1(0|\mu=m_b)=0.32\pm 0.06.
\end{eqnarray}
Fig.~\ref{Plots:FT}(a) then yields the corresponding allowed range
$\lambda_{B_S}(m_b)=(0.62\pm 0.10)$ GeV and from Fig.~\ref{Plots:FT}(b)
one obtains the estimate for the $B_s\to\gamma$ form factors
$F^{(s)}_{V}(0,0)=0.13 \pm 0.01$ and $F^{(s)}_{TV}(0,0|m_b)=0.105 \pm 0.015$.
These values are in agreement with the estimate obtained in \cite{mnk2018},
$F^{(s)}_{V}(0,0)=0.1$ and $F^{(s)}_{TV}(0,0|m_b)=0.11$, and consistent also with the recent 
lattice QCD calculations of  \cite{lattice2024}. 

Making use of the results for $R(E_\gamma=M_{B_s}/2,\mu)$ from Appendix \ref{AppendixB}
(Fig.~\ref{Plots:lambdavsmu}(a)) and the property that $R(E_\gamma=M_{B_s}/2,\mu)/\lambda_{B_s}(\mu)$ 
is scale-independent, 
we obtain the $\mu$-dependence of $\lambda_{B_s}(\mu)$ shown in Fig.~\ref{Plots:lambdavsmu}.
Evolving the allowed range of $\lambda_{B_s}(\mu)$ from $\mu=m_b$ to $\mu\simeq 1$ GeV, we obtain the estimate 
\begin{eqnarray}
\lambda_{B_s}(1\mbox{ GeV})  = 0.49\pm 0.07 \mbox{ GeV}.
\end{eqnarray}

\section{Discussion and conclusions}
A. We have formulated an efficient algorithm of obtaining the form factors $F_{V,TV}(q^2,q'^2)$ in a broad range
of momentum transfers including the region of meson resonances where the form factors cannot be calculated
using finite-order QCD diagrams. Our algorithm is based on the following steps: 

\noindent
i. We calculate the form factors $F_{V,TV}(q^2,q'^2)$ at the lowest order in $\alpha_s$ 
via the 2DAs of the $B_s$ meson in a broad range of $q^2$ and $q'^2$ far below quark thresholds where the calculation
based on finite order Feynman diagrams may be trusted. In this way we obtain, in particular, 
the dependence of $F_{V,TV}(q^2,q'^2)$ on the parameter $\lambda_{B_s}$ making use of the lowest order in $\alpha_s$.

\noindent
ii. We interpolate the obtained results by an analytic formula based on dispersion representation
for $F_{V,TV}(q^2,q'^2)$ in $q'^2$ with one subtraction. The considered analytic formula takes into account
the presence of the $\phi$-meson pole in the physical form factor and takes into account contribution of
the excited hadron states via an effective pole whose mass and coupling are fitting parameters.

\noindent
iii. We obtain predictions for the form factors $F_{V}(q^2,q'^2)$ and $F_{TV}(q^2,q'^2)$ in the physical region
by using the fitting formula in a broad range of momentum transfers including the resonance region.

\vspace{0.5cm}
B. We applied this algorithm and obtained the $B_s\to\phi$ form factors $V(q^2)$ ($T_1(q^2)$) as a residue of $F_{V}(q^2,q'^2)$
($F_{TV}(q^2,q'^2)$) of the pole located at $q'^2=M_\phi^2$ without using QCD sum
rules. In this way we avoid the systematic uncertainties related to the method of QCD sum rules 
which are not possible to control in a rigorous way
\cite{lms2007a,lms2007b,lms2009a,lms2009b,m2009}.\footnote{
We are not saying that our method avoids systematic uncertainties:
e.g., the residue in the pole has sensitivity to the ranges of values of $q^2$ and $q'^2$
where the interpolation is performed.
However, taking into account the precise knowledge of the pole location (i.e.~the physical
masses of the lightest vector mesons) reduces this uncertainty considerably.}

\vspace{0.5cm}
C. Our analysis is based on the correlation function calculated at leading order in $\alpha_s$.
Making use of the fact that the radiative corrections for the form factor $F_{V}(q^2,q'^2=0)$ 
evaluated at the scale $\mu=m_b$ turn out to be small in a broad range of photon energies $E_\gamma$, we have given arguments
that the same property is valid for the form factors $F_{V}(q^2,q'^2)$ in a broad range of values of $q^2$ and $q'^2$ far below the quark thresholds,
where our fitting procedure is carried out. In the case of the form factor $F_{TV}(q^2,q'^2)$ the radiative corrections are
not known at all, but also in this case we may conjecture that the same property is valid.  

\vspace{0.2cm}

We emphasize that it is not necessary that the radiative corrections vanish precisely at the scale $\mu=m_b$;
it is sufficient that they vanish at some scale not far from $m_b$. Then, the obtained estimate for
$\lambda_{B_s}(\mu=m_b)$ would change only marginally as the dependence of $\lambda_{B_s}(\mu)$ in the region $\mu \sim m_b$
is relatively flat. But, obviously, for putting our analysis on a solid theoretical basis, a direct evaluation of
the radiative corrections for the form factors of the tensor currents would be crucial.

\vspace{0.2cm}
{Strictly speaking, the radiative corrections in $F(q^2,q'^2)$ are of course different from those in 
$F(q^2,0)$. However, taking into account the arguments given in Section 2.C, we believe that one can draw a reasonable
conclusion about the magnitude of the radiative corrections in $F(q^2,q'^2)$ in the interesting range
of $q^2$ and $q'^2$ based on the results for the radiative corrections in $F(q^2,0)$.
Let us also emphasize that we rather need a qualitative result on the smallness of the radiative corrections
in $F(q^2,q'^2)$ in the interesting range of $q^2$ and $q'^2$.
Even if the actual size of these corrections in $F(q^2,q'^2)$ is a factor of 2-3 larger than the estimate
obtained on the basis of the known radiative corrections in
$F(q^2,0)$, this will not invalidate our results. To get a better control over
the radiative corrections in $F(q^2,q'^2)$ a calculation of such corrections would be plausible,
which is however a very formidable task.} 

\vspace{0.2cm}
Given the property that the radiative corrections are small at $\mu=m_b$, our leading-order expressions for the form factors
evaluated via the 2DAs at the scale $\mu=m_b$, containing $\lambda_{B_s}(m_b)$,
should provide a good approximation for the form factors $F_{V}(q^2,q'^2)$ and $V(q^2)$ as well as for
$F_{TV}(q^2,q'^2|\mu=m_b)$ and $T_1(q^2|\mu=m_b)$. 

Then, having at hand the dependence of $T_1(0)$ and $V(0)$ on $\lambda_{B_s}(m_b)$ and making use of the theoretical results for
$B_s\to\phi$ form factors $V(0)$ and $T_1(0|\mu=m_b)$, obtained by approaches which do not use the 2DAs of $B_s$-meson,
we extract the following allowed range s
\begin{eqnarray}
  &&\lambda_{B_s}(\mu\simeq m_b) = 0.62\pm 0.10 \mbox{ GeV} ~ , ~\\[2mm]
  &&\lambda_{B_s}(1\mbox{ GeV})  = 0.49\pm 0.07 \mbox{ GeV}.   
\end{eqnarray}
The latter result agrees nicely with thea recent estimate $\lambda_{B_s}(1 \mbox{ GeV})=0.48\pm 0.09$ GeV 
obtained using QCD sum rules in \cite{lambdabs2024}. 

Closing this paper, we believe that the proposed method for extracting the meson-to-meson transition form factor
as a residue of a dominant pole, like  $T_1(q^2)$ in the form factor $F_{TV}(q^2,q'^2)$, may be promising for calculating
other meson transition form factors and may be used as a complementary method to QCD sum rules. 

\acknowledgments
We are grateful to Matthias Neubert, Sergio Leal-Gomez, and Giuseppe Gagliardi for interesting discussions
and valuable comments.
The research was carried out within the framework of the program ``Particle Physics and Cosmology'' 
of the National Center for Physics and Mathematics.

\appendix
\renewcommand\theequation{A.\arabic{equation}}
\section{Form factor $T_1(q^2)$ for $B_s\to\phi$ from QCD sum rule \label{AppendixA}}
Here we would like to recall briefly a sum-rule calculation of the form factor $T_1(q^2)$.
The starting point is the same correlation function (\ref{tensor_penguin}) and the form factor $F_{TV}(q^2,q'^2)$. 
Eq.~(\ref{FT}) based on the calculation of QCD diagrams,
may be identically rewritten in the form of a dispersion representation in $q'^2$:
\begin{eqnarray}
\label{Eq6}
F_{TV}(q^2,q'^2)=\int\limits_0^{2\omega_0} d \xi \frac{ \phi_{B_s}(\xi)}{1-\xi}
\int ds'\frac{1}{s'-q'^2}\delta\left(s'-\frac{m_s^2+\xi(1-\xi) M_{B_s}^2-q^2\xi}{1-\xi}\right), 
\end{eqnarray}
where we introduce a short-hand notation
\begin{eqnarray}
\phi_{B_s}(\xi)\equiv -Q_s M_{B_s} f_{B_s} \left[ \phi_+(\xi)(1-\xi) +\bar\Phi(\xi)\right]. 
\end{eqnarray}
We can now make a subtraction in this spectral representation and write:
\begin{eqnarray}
\label{Eq7} 
F_{TV}(q^2,q'^2)=F_{TV}(q^2,0)+q'^2\int\limits_0^{2\omega_0} d\xi \frac{ \phi_{B_s}(\xi)}{1-\xi}
\int ds'\frac{1}{s'(s'-q'^2)}\delta\left(s'-\frac{m_s^2+\xi(1-\xi) M_{B_s}^2-q^2\xi}{1-\xi}\right).
\end{eqnarray}
Obviously, the representation (\ref{Eq6}) may not be used in the region of $q'^2$ close to the quark threshold at $4m_s^2$. 
On the other hand, from the general property of the correlation function, $F_{TV}(q^2,q'^2)$
contains a pole at $q'^2=M_\phi^2$.
\begin{figure}[b!]
\begin{center}
\begin{tabular}{c}   
\includegraphics[height=5cm]{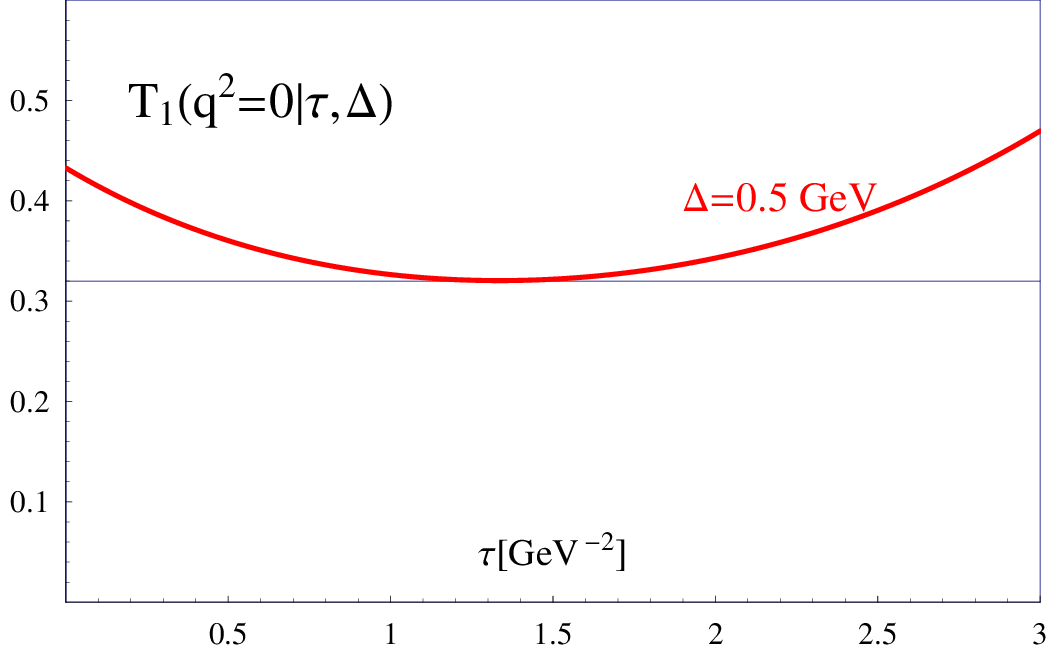} 
\end{tabular}
\caption{\label{Plots:FTSR}
The dual form factor $T_1(0|\tau,\Delta)$, a QCD sum rule estimate for $T_1(0)$; $\Delta=0.5$ GeV.
\label{Plot:T1}}
\end{center}
\end{figure}
 
We can now equate to each other two different representations for $F_{TV}(q^2,q'^2)$ --
the phenomenological side of a QCD sum rule (in the language of hadron states)
and the theoretical side (based on QCD diagrams): 
\begin{eqnarray}
\label{Eq8}
F_{TV}(q^2,q'^2)&=&R_\phi T_1(q^2)\frac{1}{M_\phi^2-q'^2}+\ldots\nonumber\\
          &=&F_{TV}(q^2,0)+q'^2\int\limits_0^1 d \xi \frac{ \phi_{B_s}(\xi)}{1-\xi}
\int ds'\frac{1}{s'(s'-q'^2)}\delta\left(s'-\frac{m_s^2+\xi(1-\xi) M_{B_s}^2-q^2\xi}{1-\xi}\right), 
\end{eqnarray}
where $\dots$ denote the contribution of hadron continuum and the excited resonances. 
To isolate the pole at $q'^2=M_\phi^2$ we perform two standard steps adopted in QCD sum rules:
\begin{itemize}
\item[(i)]
Introduce the effective threshold $s_{\rm eff}$ and apply the duality cut $s'<s_{\rm eff}$. 
We shall make use of the standard hypothesis (``Braun prescription'') based on the properties of hadron spectrum: 
\begin{eqnarray}
  \label{Eq9}
s_{\rm eff}=(M_\phi+\Delta)^2,\quad \Delta\simeq 500\;{\rm MeV}. 
\end{eqnarray}
\item[(ii)]
 Perform Borel transform $q'^2\to \tau$ $(\equiv 1/M_{\rm Borel}^2)$:
 \begin{eqnarray}
    \label{Eq10}
    1/(s'-q'^2)\to \exp(-s'\tau),\quad 1/(M_\phi^2-q'^2)\to \exp(-M_\phi^2\tau). 
\end{eqnarray}
  This Borel transform suppresses the contributions of higher-mass resonances on the phenomenological side of the QCD
  sum rule for $F_{TV}(q^2,q'^2)$ and kills all polynomial terms in $q'^2$ on the OPE side of the QCD sum rule,
  including possible subtraction terms.
\end{itemize}
Using the $\delta$-function to take the $s'-$integral, we obtain the dual form factor $T_1(q^2|\tau,\Delta)$
from which one should obtain a sum-rule estimate for the physical form factor $T_1(0)$: 
\begin{eqnarray}
\label{Eq12}
T_1(q^2|\tau,\Delta)=\frac{1}{R_{\phi}}
\int\limits_0 d \xi \frac{\phi_{B_s}(\xi)}{1-\xi}
\exp\left[\tau\left(M_\phi^2-\frac{m_s^2+\xi(1-\xi) M_{B_s}^2-q^2\xi}{1-\xi}\right)\right]
\theta\left(\frac{m_s^2+\xi(1-\xi) M_{B_s}^2-q^2\xi}{1-\xi}\le s_{\rm eff}\right).
\end{eqnarray}
Note that the actual upper limit of the $\xi$-integration in Eq.~(\ref{Eq12})
is determined by the $\theta$-function (and not by the parameter $2\omega_0$ expected to be of the order of $\sim 0.5$)
leaving the $\xi$-integration region in the dual form factor closer to zero, i.e.~$\xi\le s_{\rm eff}/M_{B_s}^2 \sim 0.1$.

The dual form factor contains two auxiliary parameters, the effective threshold 
$\Delta$ and the Borel parameter $\tau$. We shall not try to play the game with stability
windows but just fix $\Delta$ as mentioned above and take the minimal value in the Borel window
as the SR estimate for the form factor $T_1(0)$. This gives $T_1(0)\simeq 0.32$ (see Fig.~\ref{Plots:FTSR}). 

\renewcommand\theequation{B.\arabic{equation}}
\section{Radiative corrections to the form factor $F_V$ \label{AppendixB}}
The radiative corrections for the tensor current form factors $F_{TV,TA}$ are not known.
We therefore refer to the known radiative corrections for the form factors $F_{A,V}$ \cite{beneke2011}.
The necessary formulas from \cite{beneke2011} are reproduced here for the convenience of the reader. 

To the leading order in $1/m_b$, the form factors $F_{A,V}$ have the following behaviour: 
\begin{eqnarray}
  \label{B3}
F_{A,V}(E_\gamma) &=& \frac{Q_u M_{B_s} f_{B_s}}{2 E_\gamma\lambda_{B_s}(\mu)} \,R(E_\gamma,\mu)+\dots 
\end{eqnarray}
{In Eq.~(\ref{B3}), $f_{B_s}$ is the physical decay constant of the $B_s$-meson in QCD, and since
the l.h.s. does not depend on $\mu$, the ratio $R(E_\gamma,\mu)/\lambda_{B_s}(\mu)$ is also $\mu$-independent.}

The factor $R(E_\gamma,\mu)$ describing the radiative corrections can be written as a product of several factors, 
\begin{equation} 
\label{rfact} 
R(E_\gamma,\mu) = C(E_\gamma,\mu_{h1})K^{-1}(\mu_{h2})\times  
U(E_\gamma,\mu_{h1},\mu_{h2},\mu) \times J(E_\gamma,\mu), 
\end{equation} 
which come from the different scales contributing to the process, namely:

\noindent
$\bullet$
{the factor that emerges due to the replacement of the scale-dependent decay constant in HQET by 
the physical decay constant of the $B_s$-meson in QCD:}
\begin{equation} 
\label{FK} 
K(\mu) = 1 + \frac{\alpha_s C_F}{4\pi}  
\left( \frac32\ln\frac{m_b^2}{\mu^2}-2 \right). 
\end{equation}
$\bullet$ 
the factor related to the matching between the weak currents in QCD and SCET 
\begin{equation} 
\label{FC} 
C(E_\gamma,\mu) = 1 +  
\frac{\alpha_s C_F}{4\pi}\left(-2 \ln^2 \frac{2 E_\gamma}{\mu}  
+ 5 \ln \frac{2 E_\gamma}{\mu}  -\frac{3-2 x}{1-x} \ln 
  x - 2 {\rm Li}_2(1-x) -6 -\frac{\pi^2}{12} \right), \quad x=2 E_\gamma/m_b. 
\end{equation} 
\noindent
$\bullet$ the hard-collinear radiative correction 
\begin{eqnarray} 
\label{FJ} 
J(E_\gamma,\mu) = 1 +  
\frac{\alpha_s C_F}{4\pi}\left( 
\ln^2\frac{2 E_\gamma \mu_0}{\mu^2}  
-2 \sigma_1(\mu,\mu_0)  \ln\frac{2 E_\gamma \mu_0}{\mu^2} 
-1 -\frac{\pi^2}{6} + \sigma_2(\mu,\mu_0) \right) 
\end{eqnarray}
with the inverse-logarithmic moments defined as  
\begin{eqnarray} 
\frac{1}{\lambda_{B_s}(\mu)} = \int_0^\infty\frac{d\omega}{\omega}\, 
\phi_{+}(\omega,\mu), 
\qquad 
\sigma_n(\mu,\mu_0) = \lambda_{B_s}(\mu)\int_0^\infty\frac{d\omega}{\omega}\, 
\ln^n\frac{\mu_0}{\omega}\,\phi_{+}(\omega,\mu) 
\end{eqnarray} 
where $\mu_0=1\mbox{ GeV}$ is a fixed reference scale which is part of the definition of the inverse-logarithmic 
moments adopted in \cite{beneke2011}. Parameter $\omega$ in (\ref{FJ}) is related to dimensionless parameter $\xi$ in 
(\ref{2DAsa}) as $\omega=M_{B_s}\xi$. 
\vspace{0.5cm}

In \cite{beneke2011} the authors notice the absence of a common scale
$\mu$ that avoids parametrically large logarithms of order $\ln m_b/\mu_0$ in (\ref{FK}),  (\ref{FC}) and  (\ref{FJ})
and introduce the evolution factor $U(E_\gamma,\mu_{h1},\mu_{h2},\mu)$ into (\ref{rfact}).  
Its explicit expression is given in the Appendix of \cite{beneke2011}.
 
Our observation based on the above formulas and on the explicit form of the 2DAs is however a bit 
different from the conclusion of \cite{beneke2011}.
Indeed, for the 2DA of the LD model (\ref{2DAsa}), the dependence of $J(E_\gamma,\mu)$ in (\ref{FJ}) on $\mu_0$
drops out. We have checked that this property holds as well also for a different 2DA model referred to as Model IIA
in \cite{braun2017}. Thus, we expect that, at least at order ${\cal{O}}(\alpha_s)$, the factor $J(E_\gamma,\mu)$ of $\mu_0$ 
is independent of $\mu_0$. 
 
Because of this property, one can set $\mu_0$ equal to the hard scale $\mu$ thus
avoiding large logs in all three factors (\ref{FK}), (\ref{FC}) and  (\ref{FJ}). In this case, the expressions
(2.13) and (2.14) from \cite{beneke2011} and the expressions (42) and (43) from \cite{braun2004} (correspond to
setting $\mu_0\to\mu$) lead to the same results. If so, one can choose the common hard scale in all factors
(\ref{FK}),  (\ref{FC}) and  (\ref{FJ}), and the evolution factor $U(E_\gamma,\mu_{h1},\mu_{h2},\mu)$ reduces to unity. 

Moreover, at the scale $\mu=m_b$, the factor $R(E_\gamma,\mu=m_b)$ is very close to unity in a broad range of values of $q^2$.
Therefore, for $\mu=m_b$ the radiative corrections may be neglected and the leading-order analysis is expected to provide reliable results.


\begin{thebibliography}{100}
\bibitem{korchemsky2000}
G.~P.~Korchemsky, D.~Pirjol, and T.-M.~Yan, 
{\it Radiative leptonic decays of B mesons in QCD},
Phys.~Rev.~{\bf D61}, 114510 (2000).
\bibitem{braun2017}
V.~Braun, Y.~Ji, and A.~Manashov,
{\it Higher-twist B-meson Distribution Amplitudes in HQET}",
JHEP {\bf 05}, 022 (2017).
\bibitem{beneke2020}
M.~Beneke, C.~Bobeth and Y.-M.~Wang,
{\it $B_{d,s}\to\gamma l^+l^-$ decay with an energetic photon},
JHEP {\bf 12}, 148 (2020).
\bibitem{bbm2023}
I.~Belov, A.~Berezhnoy, and D.~Melikhov,
{\it Charming-loop contributions in $B_s\to\gamma\gamma$ decays},
Phys.~Rev.~{\bf D108}, 094022 (2023). 
\bibitem{bbm2024}
I.~Belov, A.~Berezhnoy, and D.~Melikhov,
{\it Nonfactorizable charming-loop contribution to FCNC $B_s\to\gamma ll$ decay},
Phys.~Rev.~{\bf D109}, 114012 (2024). 
\bibitem{wang2023}
Y.-K.~Huang, Y.~Ji, Y.-L.~Shen, C.~Wang, Y.-M.~Wang, X.-C.~Zhao, 
{\it Renormalization-Group Evolution for the Bottom-Meson Soft Function}, 
e-Print: 2312.15439 [hep-ph]. 
\bibitem{kou}
P.~Ball and E.~Kou, 
{\it $B\to \gamma e\nu$ transitions from QCD sum rules on the light cone}, 
JHEP {\bf 0304}, 029 (2003). 
\bibitem{braun2004}
V.~M. Braun, D.~Yu. Ivanov, and G.~P. Korchemsky.
{\it The B meson distribution amplitude in QCD}, 
Phys.~Rev.~{\bf D69} 034014 (2004).
\bibitem{beneke2011}
M.~Beneke and J.~Rohrwild,
{\it B meson distribution amplitude from $B\to \gamma l\nu$}, 
Eur.~Phys.~J. {\bf C71}, 1818 (2011).
\bibitem{wang2016}
Y.~M.~Wang,
{\it Factorization and dispersion relations for radiative leptonic $B$ decay}, 
JHEP \textbf{09}, 159 (2016).
\bibitem{zwicky2021}
T.~Janowski, B.~Pullin, and R.~Zwicky, 
{\it Charged and neutral $\bar B_{u,d,s}\to \gamma$ form factors
from light cone sum rules at NLO},
JHEP {\bf 12}, 008 (2021).
\bibitem{im2022}
M.~A.~Ivanov and D.~Melikhov,
{\it Theoretical analysis of the leptonic decay $B\to lll'\nu'$}, 
Phys.~Rev.~{\bf  D105}, 014028 (2022); Phys.~Rev. {\bf D106}, 119901(E) (2022).
\bibitem{khodjamirian2020}
A.~Khodjamirian, R.~Mandal, and T.~Mannel,
{\it Inverse moment of the $B_{s}$-meson distribution amplitude from QCD sum rule}, 
JHEP {\bf 10}, 043 (2020).
\bibitem{mk2003} 
F.~Kruger and D.~Melikhov,
{\it Gauge invariance and form-factors for the decay $B\to \gamma l^+ l^-$}, 
Phys.~Rev.~{\bf D67}, 034002 (2003).
\bibitem{mnk2018}
A.~Kozachuk, D.~Melikhov, and N.~Nikitin, 
{\it Rare FCNC radiative leptonic $B_{s,d}\to \gamma l^+l^-$ decays in the Standard Model}, 
Phys.~Rev.~{\bf D97}, 053007 (2018).
\bibitem{mnp2004}
D. Melikhov, O. Nachtmann, V. Nikonov, and T. Paulus,
{\it Masses and couplings of vector mesons from the pion electromagnetic, weak, and $\pi\gamma$
  transition form-factors}, Eur.~Phys.~J.~{\bf C34}, 345 (2004).
\bibitem{braun1994}
V.~M.~Braun and I.~Halperin,
{\it Soft contribution to the pion form-factor from light cone QCD sum rules},
Phys.~Lett.~{\bf B328}, 457 (1994).
\bibitem{offen2007}
A.~Khodjamirian, T.~Mannel, and N.~Offen,
{\it Form-factors from light-cone sum rules with B-meson distribution amplitudes},
Phys.~Rev.~{\bf D75}, 054013 (2007).
\bibitem{ims2020}
M.~Ivanov, D.~Melikhov, and S.~Simula, 
{\it Form factors for $B\to j_1 j_2$ decays into two currents in QCD}, 
Phys.~Rev.~{\bf  D101}, 094022 (2020).
\bibitem{ms2000}
D.~Melikhov and B.~Stech, 
{\it Weak form-factors for heavy meson decays: an update},
Phys.~Rev.~{\bf D62}, 014006 (2000).
\bibitem{ballbraun1998}
P.~Ball and V.~M.~Braun, 
{\it Exclusive semileptonic and rare B meson decays in QCD}, 
Phys.~Rev.~{\bf  D58}, 094016 (1998).
\bibitem{ballzwicky2005}
P.~Ball and R.~Zwicky,
{\it $B_{d,s}\to \rho, omega,K^*,\phi$ decay form factors from light-cone sum rules reexamined}, 
Phys.~Rev.~{\bf  D71}, 014029 (2005).
\bibitem{lattice2014}
R.~Horgan, Z.~Liu, S.~Meinel, and M.~Wingate,
{\it Lattice QCD calculation of form factors describing the rare
  decays $B \to K^* \ell^+ \ell^-$ and $B_s \to \phi \ell^+ \ell^-$},
Phys.~Rev.~{\bf D89}, 094501 (2014).
\bibitem{ivanov2016}
S.~Dubni\v{c}ka, A.~Z.~Dubni\v{c}kov\'a, A.~Issadykov, M.~A.~Ivanov, A.~Liptaj and S.~K.~Sakhiyev,
{\it Decay $B_s\to \phi \ell^+ \ell^-$ in covariant quark model}, 
Phys.~Rev.~{\bf D93}, 094022 (2016). 
\bibitem{lattice2024}
R.~Frezzotti, G.~Gagliardi, V.~Lubicz, G.~Martinelli, C.~T. Sachrajda et al, 
{\it The $B_s\to \mu^+\mu^-\gamma$ decay rate at large $q^2$ from lattice QCD}, 
e-Print: 2402.03262 [hep-lat]
\bibitem{lms2007a}
W.~Lucha, D.~Melikhov, and S.Simula, 
{\it Systematic uncertainties of hadron parameters obtained with QCD sum rules},
Phys.~Rev.~{\bf D76}, 036002 (2007).
\bibitem{lms2007b}
W.~Lucha, D.~Melikhov, and S.~Simula, 
{\it Can one control systematic errors of QCD sum rule predictions for bound states?}, 
Phys.~Lett.~{\bf B657}, 148 (2007).
\bibitem{lms2009a}
W.~Lucha, D.~Melikhov, H.~Sazdjian, and S.~Simula, 
{\it Effective continuum threshold for vacuum-to-bound-state correlators}, 
Phys.~Rev.~{\bf D80}, 114028 (2009).
\bibitem{lms2009b}
W.~Lucha, D.~Melikhov, and S.~Simula
{\it Accuracy of bound-state form factors extracted from dispersive sum rules}, 
Phys.~Lett.~{\bf B671}, 445 (2009).
\bibitem{m2009} 
D.~Melikhov, 
{\it Hadron form factors from sum rules for vacuum-to-hadron correlators}, 
Phys.~Lett.~{\bf B671}, 450 (2009).
\bibitem{lambdabs2024}
R.~Mandal, P.~S~Patil, and I.~Ray, 
{\it Probing the inverse moment of $B_s$-meson distribution amplitude via $B_s\to\eta_s$
  form factors}, arXiv:2402.16737.
\end{thebibliography}
\end{document}